# Properties of magnetite nanoparticles synthesized through a novel chemical route

**Deepa Thapa, V. R. Palkar[*], M. B. Kurup and S. K. Malik**

Tata Institute of Fundamental Research, Mumbai 400 005, India

**Abstract**

We have developed a simple precipitation route to synthesize magnetite ($Fe_3O_4$) nano-particles with controlled size without any requirement of calcination step at high temperatures. The study of these nano-particles indicates an enhancement in saturation magnetization with reduction in size down to ~10 nm beyond which the magnetization reduces. The latter is attributed to surface effects becoming predominant as surface to core volume ratio increases. From the view -point of applications, 10 nm size of magnetite particles seems to be the optimum.

*Keywords:* Magnetic Materials, Nanomaterials

[*] corresponding author : Fax: +91-22-22804610
e-mail: palkar@tifr.res.in (V.R. Palkar)

## 1. Introduction

Synthesis of magnetite ($Fe_3O_4$) nanoparticles has long been of great interest because of their immense technological applications especially in the form of ferrofluids. A ferrofluid is a colloidal suspension of suitably coated magnetite particles in a liquid medium having unusual properties due to the simultaneous fluid mechanic effects and magnetic effects [1]. Its widespread applications are in the form of seals to protect high speed CD drives, as rotary shaft seals, for improving performance of audio speakers, in oscillation damping and position sensing [2], etc. There are also promising future biomedical applications of ferrofluids. Nanoparticles with super-paramagnetic properties have great potential to achieve such desirable properties. Various methods have been developed to synthesize $Fe_3O_4$ particles in nanometer size range [see ref. 3]. However, the magnetic properties of magnetite based nanoparticles or films highly depend upon the synthesis procedure [4-6]. Here we report a novel and simple chemical route to produce magnetite ($Fe_3O_4$) nanoparticles in a size range of 5 nm to 100 nm without calcination at high temperatures. Size effect studies conducted on these nanoparticles help to optimize the size of $Fe_3O_4$ particles suitable for ferrofluid formation and some other applications.

## 2. Synthesis and Characterization

Magnetic ferrosoferric hydroxide was precipitated by mixing solutions of ferrous chloride ($FeCl_2 \cdot 4H_2O$) and ~7 Molar ammonia solution ($NH_4OH$) at 80-90 ºC. The precipitate thus obtained was filtered and left for drying. The overnight drying of the precipitate in air at *room temperature* gave the desired $Fe_3O_4$ phase. Concentration of precursor solution and precipitation rate, are the two factors that control the particle size in this process. Concentration of ferrous chloride was, therefore, varied to obtain particles in the range of 5 – 100 nm keeping the precipitation rate constant (Table 1). In order to maintain the precipitation rate more or less same, ammonia solution was poured into ferrous chloride solution at the rate of ~1cc/sec. Unlike other techniques [3], the acidification or addition of any other reagent was not required for the formation of $Fe_3O_4$ phase. Moreover, it is remarkable that in this novel synthesis process, calcination at high temperatures is not essential for the phase formation. The present process also has merits of controlling particle size and size distribution.

**Table 1. Effect of particle size on the unit cell volume, magnetic transition temperature and saturation magnetization of $Fe_3O_4$**

| Sample | Solution Concen. (%) | Particle Size nm | Unit Cell Volume $(Å)^3$ | $T_{(M)}$ °C | $M_S$ $\mu_B$/fu |
|---|---|---|---|---|---|
| M28 | 0.025 | 6.4 | 592.7 | 465 | 1.1 |
| M32 | 0.05 | 10.8 | 587.7 | 535 | 2.6 |
| M34 | 0.60 | 37.8 | 586.5 | 566 | 2.3 |
| M37 | 3.00 | 91.4 | 586.4 | 568 | 2.0 |

The magnetite particles thus produced were then characterized by various techniques. X-ray diffractometer (XRD) was used for structural phase identification (Fig. 1). Coherently diffracting domain size ($d_{xrd}$) was calculated from the width of the XRD peak under the Scherrer approximation (which assumes the small crystallite size to be the cause of line broadening) after correcting for instrumental broadening. Scanning electron microscope (SEM) was used to find out morphology of the nano-

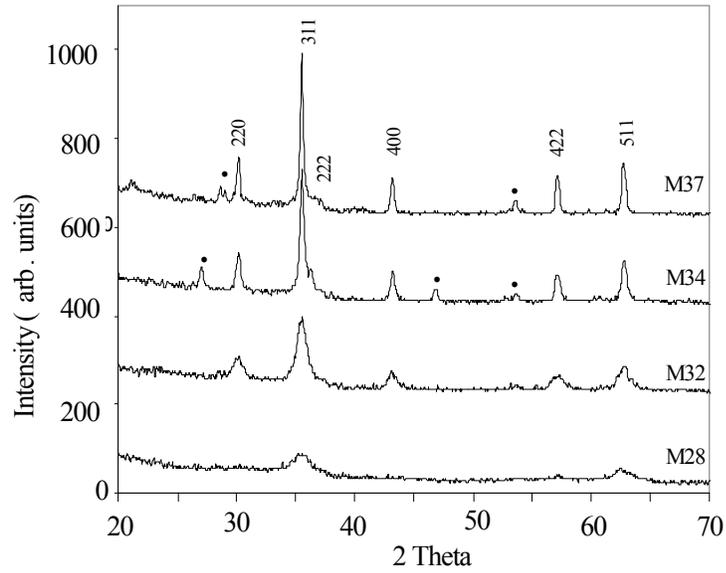

Figure 1. Indexed x-ray diffraction patterns of $Fe_3O_4$ samples with different particle sizes. '•' indicates lines from impurity phase(s).

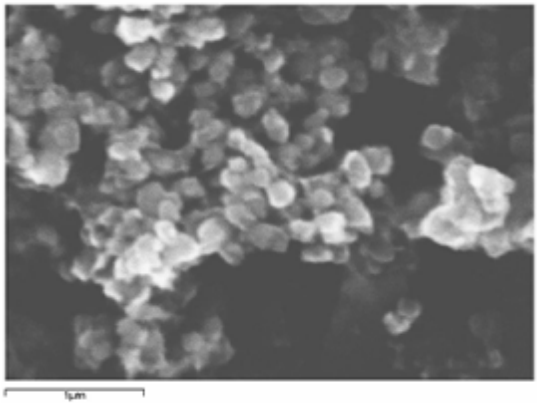

Figure 2. SEM image obtained for $Fe_3O_4$ sample (M37).

particles (Fig. 2). The images were obtained at 40 kV of accelerating voltage using 15 mm working distance and 100 nm spot size. Oxygen content in the samples was determined by using electron microprobe analyzer (EPMA). Cold pressed pellets had to be used as samples for EPMA since sintering would have caused growth in particle size. In cold pressed conditions, it was not possible to polish the surface. For accuracy in quantitative elemental analysis using EPMA, mirror finished surface is a primary requirement. Moreover, for low Z elements, such as oxygen, determination of exact number is difficult as it is. However, the purpose of carrying out electron microprobe analysis was to find out relative changes in oxygen stoichiometry with size since in

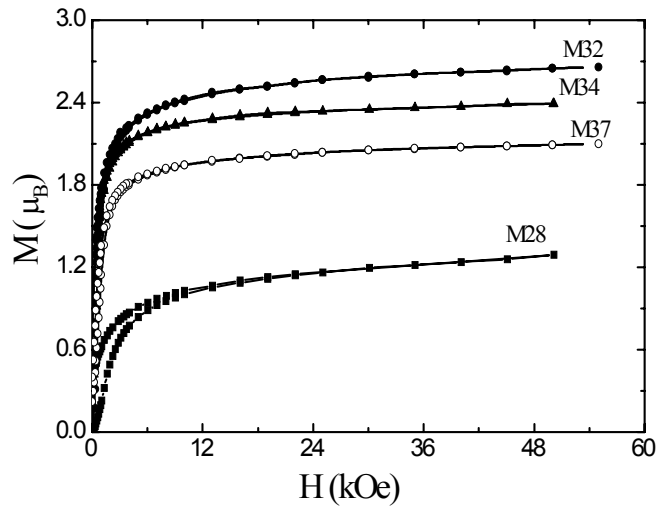

Figure 3. Magnetization (M) vs. applied field (H) isotherms at 5K obtained for different samples of $Fe_3O_4$ nanoparticles.

many other oxide systems oxygen deficiency with reduction in size has been observed [7,8]. Magnetic transition temperature ($T_M$) was determined using differential thermal analyzer (DTA). The saturation magnetization ($M_S$) was measured using a SQUID magnetometer (Table 1 and Fig. 3)

**Results and Discussion**

Figure 1 shows a typical x-ray pattern, which reveals the formation of $Fe_3O_4$ phase in all these samples. The samples M28 and M32 are single phase while M34 and M37 contain $Fe_2O_3$ as an impurity phase (~12 and ~7% respectively). As expected, the particle size decreases as the concentration of the precursor solution decreases (Table 1). SEM images indicate that the particles of all sizes are spherical in shape. Figure 2 is a representative image taken for the sample (M37) obtained by using 3% solution concentration. The expansion in unit cell volume with decrease in particle size of $Fe_3O_4$ is clear from Table 1. The reduction in particle size creates negative pressure on the lattice leading to a lattice cell volume expansion. Similar behavior has been reported earlier [9-11] for other oxide systems also. Further, the reduction in particle size also leads to decrease in magnetic transition temperature ($T_{(M)}$, Table 1.). This may be attributed to negative pressure exerted on the lattice due to unit cell volume expansion. Hedley [12] has reported similar results for fine particles of hematite. The author has attributed the depression of $T_{(M)}$ to its known sensitivity to pressure in this system. Schroer *et al* [13] have also shown that negative equivalent pressure created by size reduction could depress the transition temperature. The explanation given by Takeda *et al* [14] is based on the competing anisotropy model. According to them the dipolar anisotropy is greatly enhanced in fine particles due to large number of surface spins. The temperature at which the single–ion and dipolar terms are equal is lowered, there is also lowering of $T_{(M)}$.

The saturation magnetization values obtained by using a SQUID magnetometer (Fig. 3, Table 1.) show some interesting trends. Initially, there is an increase in saturation magnetization with decrease in particle size down to 10 nm. However, for particles less than 10 nm, there is a sudden drop in saturation magnetization value. Initial increase in magnetization value may be understood on the basis of crystal structure of $Fe_3O_4$ (given below). Also, the role played by impurity phase should not be ignored since it reduces the magnetic component in the sample.

Magnetite ($Fe_3O_4$) is a ferrimagnetic iron oxide having cubic inverse spinel structure with oxygen anions forming a fcc closed packing and iron (cations) located at the interstitial tetrahedral sites and octahedral sites. The electron can hop between $Fe^{2+}$ and $Fe^{3+}$ ions in the octahedral sites at room temperature imparting half metallic property to magnetite. The magnetic moment of the unit cell comes only from $Fe^{2+}$ ions with a magnetic moment of $4\mu_B$.

$8Fe^{3+}$ (S=5/2)

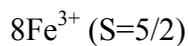 Tetrahedral sites

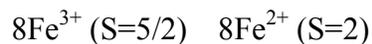 Octahedral sites

$8Fe^{3+}$ (S=5/2)   $8Fe^{2+}$ (S=2)

EPMA results suggest that, as the particle size decreases, there is a relative decrease in oxygen content of the sample, which consequently, could lead to lowering of the valance state of the cations. The observed increase in unit cell volume with reduction in particle size of $Fe_3O_4$ particles (Table 1), perhaps implies an increase in $Fe^{2+}$ content in the sample [ionic radius of $Fe^{2+}$ (0.74 Å) is larger than that of $Fe^{3+}$ (0.64 Å)]. Since the resultant magnetic moment in $Fe_3O_4$ is regarded to be due to the divalent ions ($Fe^{2+}$), the increase in magnetization with decrease in size could be justified. However, this explanation does not hold for sizes below 10 nm. Though the sample is phase pure, the drop in magnetization with further reduction in size may be attributed to surface effects [15,16]. The magnetization near the surface is generally lower than that in the interior. With increasing reduction in size, the surface effects become more predominant. In particles of the order of 4 nm, 50% of the atoms lie on the surface and, therefore, the effects are more prominent.

**Conclusions**

A novel size-controlled precipitation method is reported in this paper, which is extremely simple, cost effective and promising for synthesizing magnetite nanoparticles. An enhancement in saturation magnetization is observed with reduction in size down to ~10 nm beyond which the magnetization reduces. The latter is attributed to surface effects becoming important. Size of the order of 10 nm seems to give optimal magnetic properties and hence this size may be ideally suited for various applications. We have already synthesized a silicon oil based ferroil using these ~10 nm size particles. Studies to extend this method to prepare various substituted ferrite nano-particles are in progress.

**Acknowledgements**

The authors wish to thank Darshan Kundaliya, A.V. Gurjar, S.G. Lokhre and Ms. B.A. Chalke for expert experimental help.